\documentstyle[prb,aps]{revtex}
\begin{document}
\draft

\twocolumn[\hsize\textwidth\columnwidth\hsize\csname@twocolumnfalse%
\endcsname

\title{Indirect RKKY interaction in any dimensionality}

\author{D.N. Aristov}
\address{Petersburg Nuclear Physics Institute,
Gatchina, St. Petersburg 188350, Russia}

\date{\today}
\maketitle

\begin{abstract}
We present an analytical method which enables one to find
the exact spatial dependence of the indirect RKKY interaction
between the localized moments via the conduction electrons for
the arbitrary dimensionality $n$. The corresponding
momentum dependence of the Lindhard function is exactly found
for any $n$ as well. Demonstrating the capability of the method
we find the RKKY interaction in a system of metallic layers
weakly hybridized to each other. Along with usual $2k_F$
in-plane oscillations the RKKY interaction has the
sign-reversal character in a direction perpendicular to layers,
thus favoring the antiferromagnetic type of layers' stacking.
\end{abstract}

\pacs{
75.30.Et, 
75.40.Cx, 
75.70.-i  
}

]

The Ruderman-Kittel-Kasuya-Yosida (RKKY) interaction was found to play
an important role in various problems involving the interaction of the
localized moments in a metal via polarization of conduction electrons.
In this paper we provide the exact derivation of the spatial and
momentum dependence of RKKY interaction for arbitrary dimensionality.
In three dimensions the spatial dependence of this interaction was
obtained in Ref.\ \cite{rkky} about forty years ago. It took
some time to obtain the corresponding exact expression in two
dimensions.  \cite{rkk2d} The primary goal of this paper is thus
to present a promising analytical method for evaluation of the
corresponding expressions.

The obtained expressions being the analytical function of
dimensionality might prove to be useful for theoretical approaches
dealing with the fractional and infinitely large dimensions.
Closer to practical needs, we apply our method for an analysis of
multi-layer metal structure. We demonstrate here  the sign-reversal
character of the interaction in a direction perpendicular to layers.
The period of these oscillations coincides with a double
interlayer spacing thus favoring the antiferromagnetic ordering
of layers. It is interesting to note that the above alternation
of sign of the interaction coexists with the usual in-plane
$2k_F$ oscillations.

\section{RKKY interaction in $n$ dimensions}

We begin with conventional form of the exchange interaction
between the localized moment ${\bf J}$ and electron spin density
${\bf s}({\bf r})$ :

     \begin{equation}
     V({\bf r})=
     -A {\bf J}({\bf R}) {\bf s}({\bf r}) \delta({\bf R}-{\bf r})
     \end{equation}
Here $A$ is the exchange coupling constant.  The RKKY
interaction between two localized moments via the conducting
electrons may then be written in the following form

     \begin{equation}
     \label{rkky-def}
     H_{RKKY}= - \frac12 A^2 {\bf J}_1{\bf J}_2 \chi({\bf
     R}_{12})\ .
     \end{equation}
where $R-$dependent part of the interaction
coincides with the Fourier transform of the non-uniform static
susceptibility $\chi(q)$ (Lindhard function) and is given by :

	\begin{equation}
     \chi({\bf R}) = - T \sum_l G(i\omega_l, {\bf R} )^2
	\label{rkk-inter}
 	\end{equation}
here Matsubara frequency $ \omega_l = \pi T(2l+1)$
and the electronic Green function is

     \begin{equation}
     G(i\omega, {\bf R} ) =
     \int \frac{d^n{\bf k}}{(2\pi)^n}
     \frac{\exp(i{\bf k}{\bf R})}{i\omega -\varepsilon_k}
     \label{g-def}
     \end{equation}

We focus our attention below at the case of low
temperatures and use the limiting relation $T \sum_l \to
\int_{-\infty}^\infty d\omega/(2\pi)$.

The quadratic electron dispersion in $n$ dimensions is
explicitly assumed :

     \begin{equation} \varepsilon_{{\bf k}} = k^2/2m - \mu
     \label{en-spher}
	\end{equation}
with the Fermi energy $\mu = k_F^2/2m$.
First we use the following representation of the Green
function :

     \begin{equation}
     G(i\omega, {\bf R} ) =
     e^{-i\alpha} \!
      \int\limits_0^{\infty} \! d\tau \!
     \int \!\!\!
     \frac{d^n{\bf k}}{(2\pi)^n}
     \exp \!\left[
     i{\bf k}{\bf R} + \tau e^{i\alpha}
     \left[z -\frac{k^2}{2m} \right]
     \right]
	\end{equation}
where we introduced the value $ z = \mu + i\omega  $ and
$\alpha = sgn(\omega) \pi/2$.
The Gaussian integration over ${\bf k}$ gives
     \begin{eqnarray}
     G(i\omega, R ) &=&
     \left(\frac{m}{2\pi}\right)^{n/2}
     e^{-i\alpha(1+n/2)}
     \nonumber \\
     &\times&
     \int_0^{\infty} \frac{d\tau}{\tau^{n/2}}
     \exp[ \tau z e^{i\alpha} -
     \frac{\rho}{2\tau} e^{-i\alpha}  ]
     \label{gr-tau}
     \end{eqnarray}
with  $\rho = m R^2$.
We notice that the last integral can be expressed via
the modified Bessel (Mcdonald) function, \cite{GR} namely

	\begin{equation}
     G(i\omega, R ) = - 2
     \left(\frac{m}{2\pi}\right)^{\nu+1}
     \left(\frac{\sqrt{-2z\rho}}{\rho}\right)^{\nu}
     K_{\nu} (\sqrt{-2z\rho}),
 	\label{gr-spher}
	\end{equation}
we defined $\nu = n/2 -1 $ here.
In this equation the branch of root $\sqrt{-2z\rho}$ should be
chosen from the condition of its positive real part.  In
particular, this latter condition means that the argument of
Mcdonald function $K_\nu (\sqrt{-2z\rho}) $ has a discontinuity
at $\omega = 0$ :

     \[
     K_\nu (\sqrt{-2z\rho}) =
     \left\{
     \begin{array}{rl}
     \frac{\pi i}2 e^{\nu\pi i/2} H_\nu^{(1)}(k_FR), &
     \quad \omega \to +0 \\
     -\frac{\pi i}2 e^{-\nu\pi i/2} H_\nu^{(2)}(k_FR), &
     \quad \omega \to -0
     \end{array}
     \right.
     \]
where $H_\nu^{(1,2)}(x)$ are Hankel functions \cite{GR}.

Next we observe that one can change the variable $\omega\to Z
=\sqrt{-2z\rho}$
in (\ref{rkk-inter}) and integrate over complex $Z$ using the
exact form for the Green function (\ref{gr-spher}). Note that
the limits of integration by $Z$ are $(1\pm i)\infty$.  Without
the above discontinuity at $Z=\pm i\sqrt{2\mu\rho}$ one could
shift the integration contour to $Z\to +\infty$ and obtain zero
for (\ref{rkk-inter}) in view of the property $K_\nu(Z) \propto
e^{-Z}$.  Due to the discontinuity, the function $\chi$ has a
finite value.  After some calculations we get \cite{Ab-St}

     \begin{mathletters}
     \label{chi-spher}
	\begin{eqnarray}
     \chi(R) &=&
     \frac{m\pi}{1-n}
     \left(\frac{k_F}{2\pi R}\right)^n R^2  \Phi_n(k_FR) \\
     \Phi_n(x) &=& J_{n/2-1}(x) Y_{n/2-1}(x) + J_{n/2}(x) Y_{n/2}(x)
	\end{eqnarray}
     \end{mathletters}
This expression is the main finding of this Section. Let us
take a closer look at this result. First we note that
(\ref{chi-spher}) is the continuous function of both distance
$R$ and the dimensionality $n$.

At large distances $k_FR\gg 1$ the leading terms of the
asymptotes of Bessel functions appearing in (\ref{chi-spher})
cancel each other. Next terms produce the following
expression \cite{rkky-ani}

	\begin{equation}
     \chi(R) \simeq
     \frac{m}{k_F^2}
     \left(\frac{k_F}{2\pi R}\right)^n
     \sin( 2k_FR + \pi n /2 )
     \label{chi-spher-asym}
	\end{equation}

In particular cases of physical interest the general expression
(\ref{chi-spher}) immediately provides the exact form of RKKY
interaction in three and two dimensions \cite{rkky,rkk2d}. For
$n=3$ one has

	\begin{equation}
     \label{rkk-3}
     \chi(R) =
     -\frac{m k_F}{8\pi^3 R^3}
     \left(\cos 2k_FR - \frac{\sin 2k_FR}{2k_FR}\right),
	\end{equation}
and for $n=2$
	\begin{equation}
     \label{rkk-2}
     \chi(R) =
     -\frac{m k_F^2}{4\pi} [ J_0(k_FR) Y_0(k_FR) + J_1(k_FR)
     Y_1(k_FR) ].
     \end{equation}

The one-dimensional case can be obtained either by the
continuation \cite{Ab-St} of Bessel functions in
(\ref{chi-spher}) upon the index $n$ or by the direct evaluation
of the integral with Green function (\ref{gr-spher}) at $n=1$.
The result is \cite{rkk1d}

	\begin{equation}
     \chi(R) =  \frac{m}\pi\, {\rm si}(2k_FR) , \quad (n=1)
     \label{chi-1D}
	\end{equation}
with the sine integral
$${\rm si}(x) = \int_x^\infty \frac{dt}t \sin t$$

It is useful to define here the density of states at the Fermi level
$N(E_F)= \int d^n{\bf k}/(2\pi)^n \delta(\varepsilon_k)
= -\pi^{-1} Im G(i\omega\to +i0, R\to 0) $. From
(\ref{gr-spher}) one immediately finds
     \begin{equation}
     N(E_F) =  \frac{m}{2\pi \Gamma[n/2]}
     \left(\frac{k_F^2}{4\pi}\right)^{\frac{n}2-1}
     \label{DOS}
     \end{equation}
Now knowing the exact expression (\ref{chi-spher}) for RKKY
interaction in $R-$space one can find its correspondence in
$q-$space as follows.

     \begin{eqnarray}
     \chi(q) &=&
     \int d^n {\bf R} e^{i{\bf qR}}\chi(R)
     \\
     &=&
     \frac{m}{2(1-n)}
     \left(\frac{k_F^3}{2\pi q}\right)^{\nu}
     \int_0^\infty  dx x^{1-\nu}
     J_{\nu}\left(\frac{q}{k_f}x\right)\Phi_n(x)
     \nonumber
     \end{eqnarray}
We see that $\chi(q)$ is reduced to the Mellin convolution of
$J_\nu(x)$ and $\Phi_n(x)$. A straightforward calculation gives
then the answer expressed via the Gauss hypergeometric function
${}_2F_1[a,b,c;z]$.

     \begin{mathletters}
     \begin{eqnarray}
     \chi(q) &=&
     N(E_F)
     \phi_n\left( \frac q{2k_F} \right)
     \\
     \phi_n(x) &=& \left\{
     \begin{array}{ll}
     \displaystyle
     \frac{x^{-2}}n\,
     F\left[1,\frac12;1+\frac n2;\frac1{x^2}\right] ,& x\geq1 \\
     \displaystyle
     F\left[1,1-\frac n2;\frac32;x^2\right] ,&
     x\leq1 \end{array}
     \right.
     \label{chi-spher-q}
     \end{eqnarray}
     \end{mathletters}
Again, the result is the continuous function both in $q$ and
$n$.  From the general properties of hypergeometric function,
one has $\phi_n(0)=1$, $\phi_n(x\gg1)\sim 1/(nx^2)$ and
$\phi_n(1) = 1/(n-1)$. At last, one can easily verify
\cite{Ba-Er} that the expressions known previously
\cite{rkky,rkk2d,rkk1d} are reproduced in particular cases
$n=1,2,3$.

\section{RKKY interaction for quasi-2D case}

It is interesting to note that both exact (\ref{chi-spher}) and
asymptotic (\ref{chi-spher-asym}) expressions for $\chi(R)$ let
one mimic the ``switching on'' the extra dimensionality of a
metal by simple change of the index $n$. Thus at first sight one
could tackle the case of system of weakly hybridized metallic
planes by ascribing the dimensionality $2+\epsilon$ to it.
Actually the situation is more complicated as we discuss below.

Let us consider the (infinite) set of metallic layers, weakly
connected to each other. By this we assume the following
dispersion :

     \begin{equation}
     \varepsilon_{{\bf k}} = (k_x^2+k_y^2)/2m - \mu - \zeta \cos k_z
     \label{en-q2d}
	\end{equation}
with $\zeta \ll \mu$ and $|k_z|<\pi$ . The Fermi surface has a
cylinder-like shape with maximum and minimum in-plane radii
defined by $k_F^\pm = \sqrt{2m(\mu\pm\zeta)}$.
We write  ${\bf k}{\bf R}
= {\bf k}_\|{\bf R}_\| + k_zl$ where $l$ is the integer number
of layers.
Below we retain the definitions of $\rho= mR_\|^2$ and
$k_F=\sqrt{2m\mu}$  for the simplicity of writing.

Using (\ref{g-def}), (\ref{gr-tau}) we come to expression :

     \begin{eqnarray}
     G( R ) &=& -\frac{m}{2\pi}
     e^{-i\alpha l}
     \nonumber \\
     &\times&
     \int_0^{\infty} \frac{d\tau}{\tau}
     J_l(\tau \zeta)
     \exp[ \tau z e^{i\alpha} -
     \frac{\rho}{2\tau} e^{-i\alpha}  ]
     \label{eq-cross}
     \end{eqnarray}
The latter integral can be evaluated
( for large in-plane distances ) by the steepest descent
method. We note that when $\sqrt{2z\rho}\simeq k_FR_\| \gg 1$ ,
the principal contribution to the integral comes from the
vicinity of the point $\tau_0 = \sqrt{\rho/2z}\simeq
k_FR_\|/2\mu$, more rigorously, at $\tau =
\tau_0(1+(k_FR_\|)^{-1/2}O(1))$. It follows then that at
$k_FR_\| \lesssim (\mu/\zeta)^2$ one can replace $\tau$ by
$\tau_0$ in an argument of the Bessel function in (\ref{eq-cross}).
As a result the quasi-two-dimensional RKKY interaction is

     \begin{equation}
     \label{rkk-q2d}
     \chi(R)
     = \chi_{2D}(R_\|)\, J_l^2\left(R_\| /R_0 \right)
     (-1)^l , \quad
     R_\| \lesssim k_F R_0^2
     \end{equation}

\noindent
where $k_F R_0 = 2\mu/\zeta \gg1$ and $\chi_{2D}(R)$ is given by
(\ref{rkk-2}). Analyzing this expression we first note the
appearance of the length scale $R_0$ inversely proportional to
the strength of hybridization of layers.  Since $J_l(x) \simeq
(x/2)^l/l!$ for $x<1$, at moderate in-plane distances $R_\| <
R_0$ the interaction (\ref{rkk-q2d}) rapidly decays as a
function of $l$.

More interesting however is the fact of {\em sign-reversal
character of interaction } in a direction perpendicular to
layers. This modulation of interaction has a period {\em exactly
coinciding} with a double lattice parameter and should obviously
lead to the preferential antiferromagnetic stacking of layers.
This phenomenon is accompanied by the usual $2k_F$ oscillations
of in-plane term $\chi_{2D}(R_\|)$.

We can further clarify this point by performing the Fourier
transform with the result \cite{GR}:

     \begin{eqnarray}
     \chi({\bf k})
     &=& \int d^2{\bf R}_\|
     e^{i{\bf k}_\|{\bf R}_\|}
     \chi_{2D}(R_\|)\,
     J_0\left(2\frac{R_\|}{R_0} \cos \frac{k_z}2 \right)
     \nonumber  \\
     &\simeq& \int \frac{d\varphi}{2\pi}
     \chi_{2D}({\bf k}_\| + {\bf k}_\ast )
     \label{rkk-q2d-b}
     \end{eqnarray}
with $k_\ast = (2/R_0) \cos k_z/2 \ll 1$. The last
integration is over the angle $\varphi$ of ${\bf k}_\ast$ in the
plane, i.e. the points ${\bf k}_\| + {\bf k}_\ast $ lie on the
circle of radius $k_\ast$ and with center at ${\bf k}_\|$.

According to (\ref{rkky-def}) a maximum of $\chi_{2D}({\bf
k}_\|) $ at some ${\bf k}_0$, corresponds to the possible
in-plane magnetic ordering, characterized by this wave-vector.
We see from (\ref{rkk-q2d-b}) that the inclusion of weak
interplane hopping leads to the position of true maximum at
$k_z = \pi$, when $k_\ast =0 $. The other values of $k_z$ cause
the loss in the magnetic energy of order of the value
$\chi({\bf k}_0) k_\ast^2/k_F^2 $.

It is worth noting that the weak interplane hopping $\zeta$ and
large effective in-plane Fermi momentum $k_F$ are obviously
realized in the high-$T_c$ cuprates.  It is known that the 2D
Fermi surface  has a complicated form in these substances
\cite{ACG}.
We believe however that RKKY interaction in this
case preserves the general form (\ref{rkk-q2d}),
(\ref{rkk-q2d-b}) with anisotropic in-plane form of interaction
$\chi_{2D}({\bf k}_\|)$ and $R_0$ defined by some effective
$k_F\sim 1$.
It is also known that in the compounds
RBa$_2$Cu$_3$O$_{7-\delta}$ the subsystem of rare-earth ions
undergoes a magnetic ordering transition at low temperatures
\cite{Lynn} ; generally the type of ordering depends on a
particular ion R$^{3+}$. Remarkable fact is however that {\em
for all} substances the antiferromagnetic stacking of magnetic
R$^{3+}$ layers was reported in accordance with our finding
(\ref{rkk-q2d}).

Concluding this Section, we wish to stress the following point.
It was previously shown \cite{rkky-ani,layers} for the case of
complicated Fermi surface (FS), that the period of oscillations
( and general power-law behavior ) of the RKKY interaction is
determined by the calipering pairs of points on the FS.  These
are the points where the direction of normal to the FS is
(anti)parallel to the direction of {\bf R}. One can see that the
very notion of calipering points implies the closeness of the
Fermi surface at a given direction of {\bf R}.  On the contrast,
the FS is obviously open in our case at the $z$ direction and
the oscillations exist, albeit the roughly exponential law of
their decay.


In conclusion, we found the exact form of spatial dependence of
RKKY interaction for arbitrary dimensionality. Its counterpart
in momentum space  is also found. Applying
our method to the system of weakly hybridized metallic layers we
demonstrate the existence of spatial oscillations of indirect
RKKY exchange in the direction perpendicular to layers. The
period of oscillations equals exactly to double interlayer
spacing, which indicates the preferential antiferromagnetic
ordering of layers.

\acknowledgements
I am grateful to S.V. Maleyev for many stimulating discussions.
The partial financial support by Russian Foundation
for Basic Researches ( Grant No. 96-02-18037-a) is acknowledged.

\end{document}